\documentclass[aps,preprint,showpacs]{revtex4}
\usepackage{graphicx}  
\usepackage{dcolumn}   
\usepackage{bm}        
\usepackage{amssymb}   
\usepackage{color}
\newcommand{\pa}{\partial}
\begin{document}
\title{Rogue wave triggered at a critical frequency of a nonlinear resonant medium}
\author{Jingsong He$^{1*}$ , Shuwei Xu$^{2,3}$,  K. Porsezian$^{4}$ , Yi Cheng $^{3}$ and P. Tchofo Dinda$^5$}
\address{$^{1}$Department of Mathematics, Ningbo University, Ningbo, Zhejiang
315211, P.\ R.\ China\\
$^{2}$College of Mathematics Physics and Information Engineering, Jiaxing University,
 Jiaxing, Zhejiang 314001, P.\ R.\ China\\
$^{3}$School of Mathematical Sciences, USTC, Hefei,Anhui 230026, P.\ R.\ China\\
$^{4}$Department of Physics, Pondicherry University, Puducherry 605014, India.\\
$^{5}$ Laboratoire Interdisciplinaire Carnot de Bourgogne, UMR CNRS No. 6303, 9 Av. A. Savary, B.P. 47 870, 21078 Dijon C\'{e}dex, France}
\thanks{$^*$ Corresponding author: hejingsong@nbu.edu.cn,jshe@ustc.edu.cn}
\begin{abstract}
We consider a two-level atomic system, interacting with an electromagnetic field controlled in amplitude and frequency by a high intensity laser.
We show that the amplitude of the induced electric field,  admits an envelope profile corresponding to a breather soliton.
We demonstrate that this soliton can propagate with any frequency shift with respect to that of the control laser, except a critical frequency, at which the system undergoes a structural discontinuity that transforms the breather in a rogue wave.
A mechanism of generation of rogue waves by means of an intense laser field is thus revealed.
       \\
\noindent \textbf{KEYWORDS}: Structural discontinuity, Maxwell-Bloch equations, Rogue waves. \end{abstract}
\pacs{42.50.Md, 42.65.Tg, 02.30.Ik  }
\maketitle \mbox{\vspace{-2cm}}
\section{Introduction}
One of the most challenging aspects of modern sciences lies in the difficulty of understanding and exploitation of the effects
of the nonlinearity in numerous natural phenomena. The action of the nonlinearity has been invoked to explain the mystery surrounding many dramatic phenomena, like large amplitude waves or high intensity electric pulses observed in various fields
ranging from fluids to solid state, chemical, biological, nonlinear optical and geological systems. Nonlinear phenomena are usually modeled by evolution equations displaying a wide range of complexities. The advent of high-speed computers equipped with advanced softwares, and the development of several analytical methods, have encouraged researchers to explore this field of nonlinear phenomena, which seems for ever not saturated \cite{Book_Agrawal,Book_Scott,Book_Ablowitz,Maimbook,Kinsler1}.\\
\indent In the last five decades or so, nonlinear science has experienced phenomenal growth by the invention of fascinating new concepts like complete integrability, solitons, dromions, similaritons, rogons
  etc. Solitons have been discovered in almost all states  of matter ranging from solid medium, such as optical
  fibers, fluid dynamics, plasma systems, to Bose-Einstein condensed atomic vapor. In nonlinear optics, there
  has been intensive interest in the study of solitons in resonant optical media, including the self-induced
  transparency (SIT) in two-level atoms, normal-mode and optical simultons in three -, four-, and five-level
  media, lasing without inversion, phaseonium, photorefractive materials, photonic crystal fibre,
  electromagnetically induced transparency (EIT), and ultra-slow optical solitons.\\
\indent In addition to solitons, the
  study of rogue waves \cite{Peregrine},   boosted by the results of  \cite{Optical1},
 has  also attracted a lot of attention in nonlinear  optical system and other related areas {
  \cite{Kharif2,akhmediev1,dias1,ADegasperis,xuhewang20111jpa,hexupjpsj2012,xuhejmp2012, He2013, heg,xheheng,hewangpepre2014,heguozhangchabchoubprsa2014,Hexuchengaipadvance2015,zhangguozhouhelmp2015,qiuzhangpheprsa2015,VEZakharov,onorato,
  NatureReview2014,wrwprl2015, bbdsrprl2015}}.
  Of all collective entities that have emerged over the past three decades, rogue waves constitute undoubtedly the phenomenon the most amazing and intriguing, because of its scale, its scarcity and the particular difficulty of any direct experimental observation of this phenomenon. Yet, the current state of knowledge of this wave is still largely in its infancy, with very little experimental data available in the literature, and numerical simulations that are essentially based on statistical analysis rather than on a guidelines having a predictive characters.
In general, rogue waves always appear in a very fleeting way, as if, at first glance, they appear from nowhere and disappear without a trace \cite{nowhere}.
The specificity of rogue waves (compared to other energy localization effects), lies in the fact that there is to date no fundamental physical process
of universal character, which is recognized as being the trigger process of rogue waves.
However, previous work has revealed a variety of conditions
that may contribute to the generation of rogue waves \cite{mussot1,genty1,ankiewicz1,mahnke1,dudley1,skryabin1}.
For example,
Mussot et al conducted experiments that suggested the collisions between solitons can give rise to rogue waves \cite{mussot1}.
 Genty et al have carried out numerical simulations, which indicate that Raman effect and third-order dispersion  are phenomena
 that can trigger the formation of a rogue wave \cite{genty1}.
Other theoretical work suggest that the simultaneous action of third-order dispersion, self-steepening, and self-frequency shift, could play an important role in the generation of rogue waves in optical fibers \cite{ankiewicz1}.
More generally, the previous works have in common the fact of considering that rogue waves originate from the transformation of a collective entity
(soliton or breather) under the action of an external perturbation.
Thus, the above mentioned effects (soliton collisions \cite{mussot1},
Raman effect \cite{genty1,mahnke1}, third-order dispersion \cite{genty1,ankiewicz1,mahnke1},
self-steepening and self-frequency shift \cite{ankiewicz1}, modulational instability \cite{dudley1}, etc.),
clearly appear as external perturbations acting on a stable collective entity.
 \\

In the present work, we consider a two-level atomic system controlled by an intense laser field,
 and we show that rogue waves may also appear in this system but through
a completely different mechanism, when compared with those reported so far in the literature.
By modeling the system by the Maxwell-Bloch equations,
 we find that the amplitude of the electric field which results from the interaction between the atomic system
and the control laser,  admits an envelope profile corresponding to a breather soliton, which
 can propagate with any frequency shift with respect to that of the control laser, except  a critical frequency,
at which the system undergoes a structural discontinuity that transforms the breather in a rogue wave.
The interest of this generation process of rogue waves lies  is the prospect to generate rogue waves in a finely controlled manner.
Indeed, assuming that we can vary the soliton frequency (e.g., by a modulator coupled to the control laser, or by a frequency-shifted feedback laser),
 then two situations are possible: (i)
if the soliton frequency passes through the critical frequency without stopping, the breather  will passe fleetingly by a state  of rogue wave before returning to the breather state.
(ii) If the frequency remains fixed exactly on the critical frequency, then the
rogue wave may exist with a relatively long lifetime.

The two-level atomic system we have chosen to demonstrate this generation mechanism
of rogue waves is an excellent testing ground because it offers the unique
 opportunity to enable analytical demonstrations without resort to any statistical analysis.
\\

\section{Theoretical model and analytical results}

\subsection{Theoretical model}

We consider a resonant two-level optical medium that can be described by the following Maxwell-Bloch (MB) equations \cite{famb}:
\begin{eqnarray}
 &&E_{\xi}=\rho , \nonumber\\
 &&\rho_{\tau}=-2i\rho \eta +E N,\nonumber\\
 &&N_{\tau}=-\frac{1}{2}(E \rho^*+E^* \rho).\label{SIT}
 \end{eqnarray}
Here, $\tau$ and $\xi$ correspond respectively  to the time and space coordinates,
 $E(\tau,\xi)$ is the complex electric field envelope resulting from the interaction
between the control laser and the two-level atomic system.
 $\rho(\tau,\xi)$  is the out-of-phase and  in-phase components
 of the  induced polarization, $N(\tau,\xi)$  is the normalized population inversion, the parameter $\eta$ is the deviation
of the transition frequency of the given two-level atom, from the mean frequency.
The MB equations can be used to describe the
SIT phenomena in a resonant two-level optical medium.
Lamb \cite{famb} gave an elaborate study
 on different limiting cases of the MB equations. The connection between MB equations and the sine-Gordon equation has been reported.
 The integrability of the  MB equations is given by ref. \cite{Ablo}. In our recent work, we have considered higher order dispersion
 and nonliner effects in the form of Maxwell-Bloch and Hirota equation and obtained the breathers, bright and dark rogue wave solutions\cite{He2013}.

\subsection{Basic idea and results}

Eq.(1) has been well studied and many different kinds of solutions have been reported.As mentioned above, the rogue wave solutions of the above equation is well
documented in the literature, to avoid repitition, we are giving only the final form of the solution which is of our current interest. Though the proposed system is
well investigated,
in this paper, our prime aim is to identify  an important observation about the change in phase and group velocity bu suitably tuning the frequency of the modulated laser.
By using determinant representation of the Darboux transformation \cite{matveev,hedarbroux},
 we obtain the following breather solutions (i.e. periodic traveling wave) of the MB equations,
\begin{eqnarray}
&&{}\mbox{\hspace{-0.5cm}}E^{[1]}=\exp(i(a\xi+b\tau))(d-4\beta_1\frac{\delta_2}{\delta_1}),\\
&&{}\mbox{\hspace{-0.5cm}}\rho^{[1]}=i a d \exp(i(a\xi+b\tau))\mbox{\hspace{-0.1cm}}-\mbox{\hspace{-0.1cm}}4\beta_1\pa_{\xi} (\frac{\delta_2\exp(i(a\xi+b\tau))}{\delta_1}),  \\
&&{}\mbox{\hspace{-0.5cm}}N^{[1]}=-2a\eta-a b+4i\pa_{\xi} (\frac{\delta_3}{\delta_1}), \\
\text{where }~~&&{}\mbox{\hspace{-0.5cm}}\delta_1=r_1\cosh(M_1)+r_2\cos(M_2),\nonumber\\
&&{}\mbox{\hspace{-0.5cm}}\delta_2=r_2\cosh(M_1)+r_1\cos(M_2)+2i((-r_3+r_4)\sinh(M_1)\nonumber\\
&&{}\mbox{\hspace{-0.2cm}}+(r_5+r_6)\sin(M_2)),\nonumber\\
&&{}\mbox{\hspace{-0.5cm}}\delta_3=-2i\beta_1((r_5+r_6)\sinh(M_1)+(r_3-r_4)\sin(M_2)),\nonumber\\
&&{}\mbox{\hspace{-0.5cm}}M_1=\frac{1}{2((\alpha_1-\eta)^2+\beta_1^{2})}(R_1((-2((\alpha_1-\eta)^2+{\beta_1}^2)\tau\nonumber\\
&&{}\mbox{\hspace{0cm}}+a(\eta-\alpha_1)\xi)+a\beta_1 R_2 \xi),\nonumber\\
&&{}\mbox{\hspace{-0.5cm}}M_2=\frac{1}{2((\alpha_1-\eta)^2+\beta_1^{2})}(R_2((-2((\alpha_1-\eta)^2+{\beta_1}^2)\tau\nonumber\\
&&{}\mbox{\hspace{0cm}}+a(\eta-\alpha_1)\xi)-a\beta_1 R_1 \xi),\nonumber\\
&&{}\mbox{\hspace{-0.5cm}}r_{1}=R_{1}^2+R_{2}^2+(d+2\beta_1)^2+(b+2\alpha_1)^2,\nonumber\\
&&{}\mbox{\hspace{-0.5cm}}r_{2}=-{R_{1}}^{2}-{R_{2}}^{2}+(d+2\beta_1)^{2}+(b+2\alpha_1)^{2},\nonumber\\
&&{}\mbox{\hspace{-0.5cm}}r_{3}=R_1(b+2\alpha_1), \ r_{4}=R_2(d+2\beta_1),\nonumber\\
&&{}\mbox{\hspace{-0.5cm}}r_{5}=R_1(d+2\beta_1), \ r_{6}=R_2(b+2\alpha_1),\nonumber\\
&&{}\mbox{\hspace{-0.5cm}} \sqrt{4{\beta_1}^2-d^2-(b+2\alpha_1)^2+4i\beta_1(2\alpha_1+b)}\triangleq R_1+i R_2.\nonumber
\end{eqnarray}

By comparing with the breathers in ref. \cite{HParkr}, here $a$ is a constant (wave number),
 $b$ is the control-field frequency, $d$ is the control-field amplitude. Moreover, other constants $\alpha_1,\beta_1 $  in the above solutions originate from the Darboux transformation.

The breather solution  $E$ is plotted in Figure (1a)  with $ a=\frac{1}{\sqrt{(b+2\eta)^{2}+d^2}}$.  It is clear that this breather
propagates without any distortion. In addition, the electric field component admits bright breather type, whereas polarization and population inversion admit dark type with multiple peaks in each case {(see  Figures (1b) and (1c))}.  From the above solutions, by a simple calculation, the group velocity $\widetilde{V}_g$
and phase velocity $\widetilde{V}_p$ of $E^{[1]}$ are deduced in the following form
$\widetilde{V}_{g}=\frac{2R_1((\alpha_1-\eta)^2+{\beta_1}^{2})}{a(R_1\eta-R_1\alpha_1+R_2\beta_1)}, \widetilde{V}_{p}=-\frac{2R_2((\alpha_1-\eta)^2+{\beta_1}^{2})}{a(R_1\beta_1+R_2\alpha_1-R_2\eta)}$.

Now, a careful inspection of these formulas of $\widetilde{V}_{p}$  and $\widetilde{V}_{g}$, reveals the existence of a jump in these velocities when \textbf{$\alpha_1\mapsto -\frac{b}{2}$} and \textbf{$\beta_1 \mapsto-\frac{d}{2}$}.
To shed more light on this velocity jump,
 we  choose $ d=1, \eta=\frac{1}{2}, \alpha_1=\frac{1}{2}, \beta_1=-\frac{1}{2} $, then  $\widetilde{V}_g$
and $\widetilde{V}_p$ become two simple functions of  $b$ as
\begin{eqnarray}
&&{}\mbox{\hspace{-0.5cm}}V_{p}\mbox{\hspace{-0.1cm}}=\mbox{\hspace{-0.2cm}}-\mbox{\hspace{-0.2cm}}\frac{(2b\mbox{\hspace{-0.1cm}}
+\mbox{\hspace{-0.1cm}}1\mbox{\hspace{-0.1cm}}+\mbox{\hspace{-0.1cm}}b^2
\mbox{\hspace{-0.1cm}}+\mbox{\hspace{-0.1cm}}\sqrt{(5+b^2+2b)(b+1)^2})\sqrt{b^2+2b+2}}{2(b+1)}\mbox{\hspace{-0.1cm}}, \nonumber\\
&&{}\mbox{\hspace{-0.5cm}}V_{g}=\frac{2(b+1)\sqrt{b^2+2b+2}}{2b+1+b^2+\sqrt{(5+b^2+2b)(b+1)^2}}.
\end{eqnarray}
The plotting of the velocity as a function $b$, visible in {Fig. 2}, shows more clearly  an abrupt jump of the velocity, at $b=b_c=-1$.
It is also easy to verify this behavior by a straightforward calculation of right and left limits:
$\lim_{b\rightarrow -1^{-}} V_g=-1, \lim_{b\rightarrow -1^{+}} V_g=1$. Similarly, $V_p$ also exhibits a jump from $1$ to $-1$ at $b=b_c=-1$.
As this velocity jump occurs during the variation of the only parameter $b$
(keeping all other  parameters at constant values), without a reversal of the sign of b,
this velocity jump necessarily reflects  a structural discontinuity, localized at the point
$b_c=-1$, which we define as the {\it critical frequency} of the system.
Here, an outstanding point of the  structural discontinuity is the
reversal of the sign of the breather velocity without a reversal of the sign of b.
Thus, the soliton does not exist at the critical frequency $b=b_c$. For
$b\neq b_c$, the soliton propagates, but in respectively opposite directions
for $b < b_c$, and $b> b_c$.
However, it should be noted that reversing the sign of $b$ around the critical frequency $b_c$
does not mean that two counter-propagating beams can simultaneously exist in this system,
because the equation (1) is a first-order model, which corresponds to the unidirectional approximation
of the propagation of the light field. 
 In what follows we demonstrate below a fact even more important, that the locking
 of the system parameters on the structural discontinuity enables generation of a rogue wave.

Hence, similar to the other several integrable systems \cite{xuhewang20111jpa,hexupjpsj2012,xuhejmp2012, He2013, heg,xheheng,hewangpepre2014,heguozhangchabchoubprsa2014,Hexuchengaipadvance2015,zhangguozhouhelmp2015,qiuzhangpheprsa2015},
by doing Taylor expansion in the above breather solutions with respect to special values of parameters
 $\alpha_1\mapsto -\frac{b}{2}$ and $\beta_1 \mapsto-\frac{d}{2}$, the
 first-order rogue waves are constructed in the form
\begin{eqnarray}
&&{}\mbox{\hspace{-0.5cm}}E_r^{[1]}=\exp(i(a\xi+b\tau))(d+2d\frac{\tilde{\delta_2}}{\tilde{\delta_1}}),\\
&&{}\mbox{\hspace{-0.5cm}}\rho_r^{[1]}=i a d \exp(i(a\xi\mbox{\hspace{-0.1cm}}+\mbox{\hspace{-0.1cm}}b\tau))\mbox{\hspace{-0.1cm}}+\mbox{\hspace{-0.1cm}}2d\pa_{\xi} (\frac{\tilde{\delta_2}}{\tilde{\delta_1}}\exp(i(a\xi\mbox{\hspace{-0.1cm}}+\mbox{\hspace{-0.1cm}}b\tau))), \\
&&{}\mbox{\hspace{-0.5cm}}N_r^{[1]}=-2a\eta-a b+4i\pa_{\xi} (\frac{\tilde{\delta_3}}{\tilde{\delta_1}}),\\
\mbox{where}~~&&{}\mbox{\hspace{-.5cm}}  \ \  \frac{\tilde{\delta_2}}{\tilde{\delta_1}}=\frac{-4 d^2 S_1^2-(a d^2 \xi+i S_2)^{2}}{a^2 d^4 \xi^2+4 d^2 S_1^2+ S_2^2},\nonumber\\
&&{}\mbox{\hspace{-0.5cm}}\frac{\tilde{\delta_3}}{\tilde{\delta_1}}=\frac{-2i d^2 S_1 S_2}{a^2 d^4 \xi^2+4 d^2 S_1^2+ S_2^2}, S_2=(b+2\eta)^2+d^2,\nonumber\\
&&{}\mbox{\hspace{-0.5cm}}S_1=-\frac{1}{2}(((b+2\eta)^2+d^2)\tau-a(b+2\eta)\xi).\nonumber
\end{eqnarray}
Thus, {Fig. 3} illustrates three parameters corresponding to the rogue waves,
 namely, $b,d$,and $\eta$.

\subsection{\bf Comparison with previous work}

 It is worth comparing our results to those of Ref.\cite{HParkr}, which also studied a similar system. From the breather solution of Eq.(6) given in Ref.\cite{HParkr}, setting $S\rightarrow 0$, it reduces to a rogue wave solution.
For this solution, the group velocity of $E(X,T)$ is  found to be $-\frac{2K}{\kappa D_c}$, which is always positive. Thus this group velocity does not exhibit a jump with respect to  $\Delta_c$ and $E_0$. It is clear that in Ref. \cite{HParkr},  the effect of the discontinuity is visible rather on the S parameter, at the critical value $S=S_c=0 $.  In addition, there exists a correspondence between the results of Ref. \cite{HParkr} and our calculation: $E_0\longleftrightarrow d,\Delta_c \longleftrightarrow b$.
One of the objectives in ref. \cite{HParkr}, is the lowering of the velocity of light, whereas here our goal is rather the generation of rogue waves. Although the two objectives are clearly different, the physical mechanisms used in the two cases have common feature, to be controlled by a parameter located in the immediate vicinity of a critical value that we have identified in this study as being a structural discontinuity.
Indeed, in Ref.\cite{HParkr}  the parameter denoted $S$ temporal width of the breather is lowered in order to
 decrease the breather velocity: $S\longrightarrow 0 $. In fact, the critical value of this parameter ($S=S_c=0 $) is nothing but a structural
  discontinuity of the system. This critical value is not reached in Ref.\cite{HParkr} (because it is not
   the intention).  However, we can observe that, for $S=S_c=0$, the soliton found in Ref.\cite{HParkr}  is no longer defined, and that the system admits rather a rogue wave. Figure 4 is plotted analytically for this rogue wave but which is not mentioned in Ref.\cite{HParkr}.
     On the other hand, in both cases, the breather amplitude does not present any structural discontinuity (in the parameter regions considered). Indeed, according to our definition of $R_1$ and $R_2$ of breather solution $E^{[1]}$, there is no velocity jump if we set $\alpha_1=-\frac{b}{2}$ but retain $d$. Similarly, in figure 2
of Ref.\cite{HParkr}, it is trivial to verify that there is no structural discontinuity about the breather amplitude $E_0$.
    At last, by comparing with RW solution in Eq.(6.10) of  Ref.\cite{veev},  the first-order RW solution
 $E_{r}^{[1]}$ has a similar formulation.  However,  the explicit expressions of two other first-order RW solutions
 $\rho_{r}^{[1]}$ and $N_{r}^{[1]}$ have not been given in  Ref.\cite{veev}.

Thus, from a more fundamental viewpoint, our above results demonstrate
the existence of rogue waves triggered by a structural discontinuity of the system {at a critical frequency $b_c$.}
Consequently, the broader concept of structural discontinuity that we have introduced in our study
is a key point of the mechanism for generating rogue waves in the two-level atomic system.

\section{Conclusion}
 Though the MB equations have been well studied over the past five decades or so,
in this work, we have highlighted three important features of this system which has not
been reported so far in the literature.\\
{\bf(i)} The wave velocity can be finely adjusted by carefully tuning the breather frequency $b$.\\
{\bf(ii)} The system is endowed with a structural discontinuity that manifests itself
 (at a critical frequency $b_c$), by a reversal of the direction of propagation
when $b$ crosses $b_c$.\\
{\bf(iii)} The system admits (first-order) rogue wave solutions, but exactly
at the critical frequency.
This situation reflects the difficulty to generate rogue waves having a long lifetime.
Indeed, if the breather frequency
  passes through the discontinuity, the breather will pass
  fleetingly by a state  of rogue wave and return to breather state.
  But if the frequency can be locked exactly on the discontinuity,
  the  rogue wave may exist with a relatively long lifetime,
  which is a key requirement in the prospect of practical applications.

{\bf Acknowledgments} {\noindent \small This work is supported by the NSF of China under Grant No.11271210 and K.C.
Wong Magna Fund in Ningbo University.  J. He thanks sincerely Prof. A. S. Fokas for arranging the visit to Cambridge University in 2012-2015 and for many useful discussions.  K.P. thanks the DST, IFCPAR, NBHM and CSIR, Government of India, for the financial support through major projects. We thank referees of early versions of this paper for many useful suggestions. }

\clearpage

\setcounter{figure}{0}
\begin{figure}[!ht]\label{figbreathers}
\begin{center}
(a)\includegraphics[height=7cm]{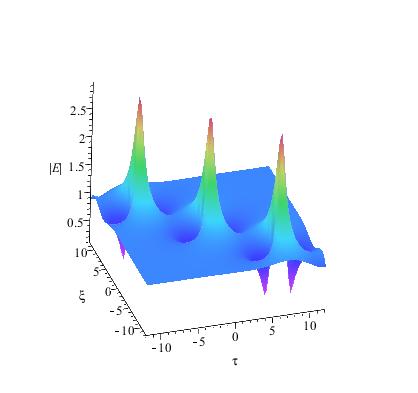}\mbox{\hspace{-1cm}} (b)\includegraphics[height=7cm]{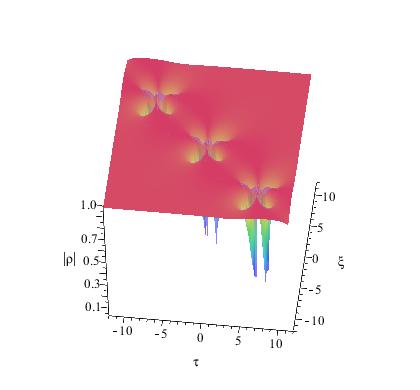}
(c)\includegraphics[height=7cm]{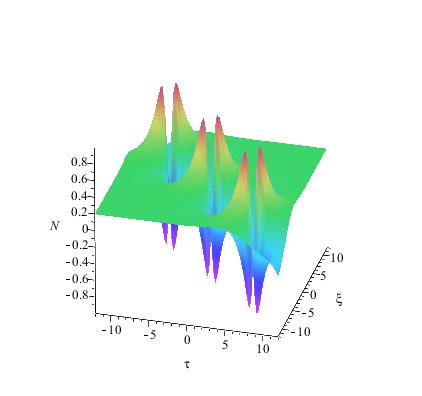}
\mbox{}\vspace{-.3cm}
 \caption{(Color online) The dynamical evolution of three breathers in the MB equation under
$d=1, \eta=\frac{1}{2}, \alpha_1=\frac{1}{2}, \beta_1=-\frac{1}{2},  b=-\frac{6}{5} $. From Figs.(a) to (c), they are $|E^{[1]}|, |\rho^{[1]}|$ and $N^{[1]}$.}
\end{center}
\end{figure}

\begin{figure}[!ht]
\begin{center}
\mbox{\hspace{-1cm}}
 (a)\includegraphics[height=6cm]{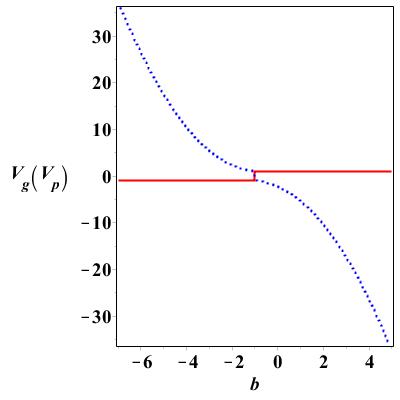}
\mbox{\hspace{0.2cm}}
(b)\includegraphics[height=6cm]{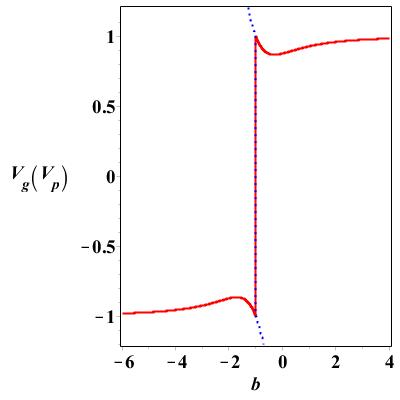}
\mbox{}\vspace{.3cm}
 \caption{(Color online)
 The structural discontinuity of the MB system.   (a) Jump in group velocity $V_g$(red,solid) and phase velocity $V_p$ (blue,dot) of the breather $E^{[1]}$ with parameters as given in Fig.1. Panel (b) is the local picture  of panel (a) around $b=-1$.}
\end{center}
\end{figure}

\begin{figure}[!ht]
\begin{center}
\mbox{\hspace{-0.5cm}}(a)\includegraphics[height=7cm]{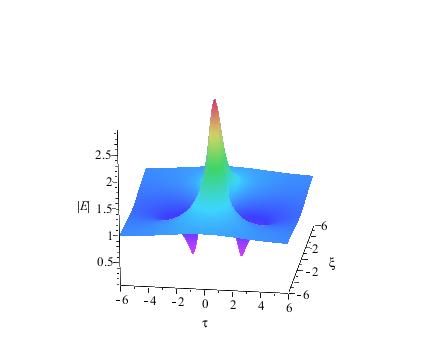}
\mbox{\hspace{-1.5cm}}(b)\includegraphics[height=7cm]{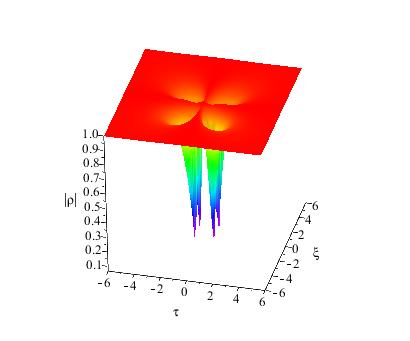}
\mbox{\hspace{-1cm}}(c)\includegraphics[height=8cm]{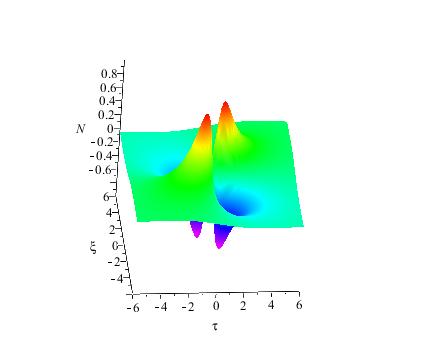}
\mbox{}\vspace{-.3cm}
 \caption{{(Color online)
Three first-order rogue waves in MB equations under $d=1, b=-1, \eta=\frac{1}{2}$.  From Figs.(a) to (c), they are $|E_r^{[1]}|, |\rho_r^{[1]}|$ and $N_r^{[1]}$.}}
\end{center}
\end{figure}

\begin{figure}[!ht]
\begin{center}
\includegraphics[height=9cm]{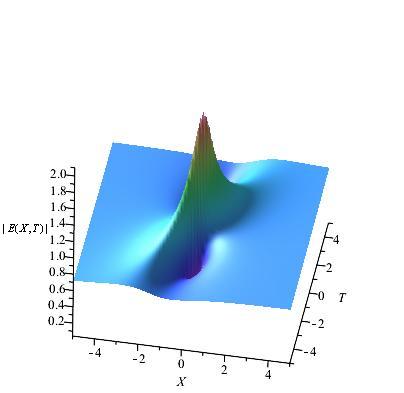}
 \caption{{(Color online)
The first-order rogue wave $|E(X,T)|$ is derived from Eq. (6) of reference \cite{HParkr} by $S\rightarrow 0$ with parameters $E_0=\frac{7}{10},\kappa=3,\Delta_c=1$. } }
\end{center}
\end{figure}

\end{document}